\documentclass[twocolumn,showpacs,preprintnumbers,amsmath,amssymb]{revtex4}

\usepackage{graphicx}% Include figure files
\usepackage{dcolumn}% Align table columns on decimal point
\usepackage{bm}% bold math
\newcommand{\comment}[1]{}
 \begin{document}

\title{Cascading Dynamics in Modular Networks}

\author{Aram~Galstyan and  Paul Cohen}
\affiliation{
Information Sciences Institute\\
University of Southern California\\
4676 Admiralty Way, Marina del Rey, CA 90292-6695\\
}

\date{\today}

\vspace{8mm}

\begin{abstract}

In this paper we study a simple cascading process in a structured heterogeneous population, namely, a network composed of two loosely coupled communities. We demonstrate that under certain conditions the cascading dynamics in such a network  has a two--tiered structure that characterizes activity spreading at different rates in the communities. We study the dynamics of the model using both simulations and an analytical approach based on annealed approximation, and obtain good agreement between the two. Our  results suggest that network modularity might have  implications in  various applications, such as epidemiology and viral marketing. 
\end{abstract}

\maketitle

\section{Introduction}
Networks are useful paradigm for studying complex systems composed of large numbers of interconnected components~\cite{albert2002,newmansiam2003,boccaletti2006}. There has been a growing interest in applying network analysis to examine  various  biological~\cite{kauffman1993,jeong2000}, ecological~\cite{montoya2002,camacho2002,jordano2003}, technological~\cite{albert1999,albert2000,kleinberg1999}, and even political~\cite{mason2005} systems. Research  on various statistical properties of such networks has revealed many interesting phenomena. For instance, the scale--free degree distributions observed in many real--world networks  have significant implications for various dynamical processes on such networks. In particular, the dynamics of SIR (susceptible--infected--removed) epidemic processes in  certain scale--free networks are characterized by vanishing threshold for epidemics~\cite{vespignani2001}, in sharp contrast with results for the random Erdos-Renyi networks. 

Another interesting property of networks is modularity, the tendency of nodes  to partition themselves into $communities$~\cite{girvan2002,M.E.J.Newman06062006}. Loosely speaking, a community is a group of nodes for which the density of links within a group is higher than across the groups. Much recent research has focused on  methods for detecting and analyzing community structure in networks (for a recent review of existing approaches see~\cite{1742-5468-2005-09-P09008} and references therein).  However, the dynamical properties of modular networks have received relatively little attention despite the potential importance of the subject to problems such as  epidemiology, viral marketing, and so on.  Consider, for instance,  word--of--mouth (or viral) marketing of a new product.  If different consumer groups have different rating criteria for the product, or different reaction to marketing strategies, then one needs to model how influence propagates within and across communities to predict whether the product will be a hit, or confined to a small subset of consumers. A similar argument holds for epidemic models where nodes are heterogeneous with respect to their susceptibilities and/or  interactions patterns~\cite{gupta1989,Aral1999,liljeros2003}. For instance, Gupta {\em et. al.} showed that the transient dynamics of sexually transmitted infection  can be very different, depending on whether the network of sexual contacts is assortative or disassortative~\cite{gupta1989}.  More recent work has addressed the role of the modularity on dynamics of cascading failures in scale free networks~\cite{wu2006}, and  synchronization patterns of networked oscillators~\cite{arenas2006}. In particular, Ref.~\cite{arenas2006} demonstrated that the modularity of  networks is strongly reflected in their synchronization dynamics, so that communities emerge as connected groups of synchronized  oscillators.

The goal of this paper is to further improve our understanding of the connection between network modularity  and its dynamics. Specifically, we examine the effects of modularity on a simple, threshold--based activation process on networks.  Starting with a modified version of Watts' cascading model~\cite{watts2002}, we study its dynamical properties for networks composed of two loosely coupled communities. Our main observation is that if the initially active nodes ({\em seeds}) are contained in one of the communities, then under certain conditions the cascading process has a two--tiered structure, that is, the peaks of the activation dynamics in each community are well separated in time.  We present results of simulations as well as analytical results based on annealed approximation, and observe a good agreement between the two.   

\section{Model}
Let us  consider a network where each node is in one of two states: passive and active. Initially, all but a small fraction of {\em seed} nodes are passive. During the cascade process, a passive node will be activated with probability that depends on the state of its neighborhs. In Watt's original model~\cite{watts2002}  this probability is $p = \Theta(h_i/k_i - \phi)$, where $\Theta$ is the step function, $h_i$ and $k_i$ are the number of active neighbors and the total number of  the neighboring nodes, respectively, and $\phi_i$ is the activation threshold for the $i$--th node.   Here we consider a slight modification of the original model by using a threshold condition  on the $number$ of active neighbors rather than their fraction: $p = \tau^{-1}\Theta(h_i - H_i)$, where $\tau$ determines the time--scale of the activation process.  For the sake of simplicity, we assume that all nodes have the same activation threshold, $H_i=H$ for all $i$. 

Clearly, the dynamics of the cascade process will depend on both network structure and the threshold parameter $H$.  Here we are interested in the case when the network is composed of two loosely coupled communities. Namely, we consider a random graph consisting of $N = N_a+ N_b$ nodes of two different type, $a$ and $b$. The probabilities of edges between nodes of different types are $\gamma_{aa}$, $\gamma_{bb}$ and $\gamma_{ab}=\gamma_{ba}$, and  the average connectivity between nodes of the respective types are then $z_{aa}=\gamma_{aa}N_a$, $z_{bb}=\gamma_{bb}N_b$, $z_{ab} =\gamma_{ab}N_b $ and $z_{ba} = \gamma_{ab}N_a$.   We want to find  out how the modularity of the network, as described by the coupling between the groups, affects the cascading process.

Let $\rho_a^0$ and $ \rho_b^0$ be the fraction of seed nodes in each population. Further, let $P_{a}(h;t)$ and $P_{b}(h;t)$ be the probability distribution that a randomly chosen node of corresponding type is connected with exactly $h$ active nodes at time $t$.  It is easy to see that $P_{a}(h;t=0)$ and $P_{b}(h;t=0)$ are Poisson distributions with means $z_{aa}\rho_a^0 + z_{ab}\rho_b^0$ and $z_{bb}\rho_b^0+z_{ba}\rho_a^0$, respectively. To study the dynamics of the process, we need to estimate these distributions for later times. This is particularly straightforward to do within the {\it annealed approximation}, e.g.,  by  ``rewiring" the network after each iteration~\cite{derrida1986}. Indeed, since all edges of corresponding type are equally likely, it is easy to see that $P_{a}(h;t)$ and $P_{b}(h;t)$ are still given by Poisson distribution, with the means that now depend on the fraction of active nodes $\rho_a(t)$ and $\rho_b(t)$: $P_{a,b}(h;t) = Poisson(\lambda_{a,b}(t))$, where $\lambda_a = z_{aa}\rho_a(t)+z_{ab}\rho_b(t)$ and $\lambda_b = z_{bb}\rho_b(t)+z_{ba}\rho_a(t)$.

On the first step of the cascading process,  the fraction of activated nodes of each type is given by $\tau^{-1}\sum_{h\ge H}P_{a,b}(h;t=0)$. In later iterations, we can calculate the fraction of active nodes as follows. Let us consider, for instance, $a$ nodes. There are  $N_a(1-\rho_a(t))$ passive nodes at time $t$, and each one of these nodes will be activated with the rate $\tau^{-1}\sum_{h\ge H}P_{a}(h;t)$. Also, due to the rewiring,  some of the $N_a(\rho_a(t) - \rho_a^0)$ active nodes will switch to passive state with the rate $\tau^{-1}\sum_{h< H}P_{a}(h;t)$. We note that {\em the seed nodes never de--activate}. Combining these together, and using the normalization condition $\sum_{h=0}^{\infty}P_{a,b}(h;t)=1$, we obtain the in the continuos time limit
\begin{equation}
\tau\frac{d\rho_{a,b}}{dt} = 1 -\rho_{a,b} -  (1 - \rho_{a,b}^0) Q(H; \lambda_{a,b})
\label{eq:dyn0}
\end{equation}
where $Q(n,x)=\sum_{k<n}e^{-x}x^k/k!$ is the regularized gamma function. 

\begin{figure}[]
\center
\begin{tabular}{c}
(a)\\
\includegraphics[ width=0.45\textwidth]{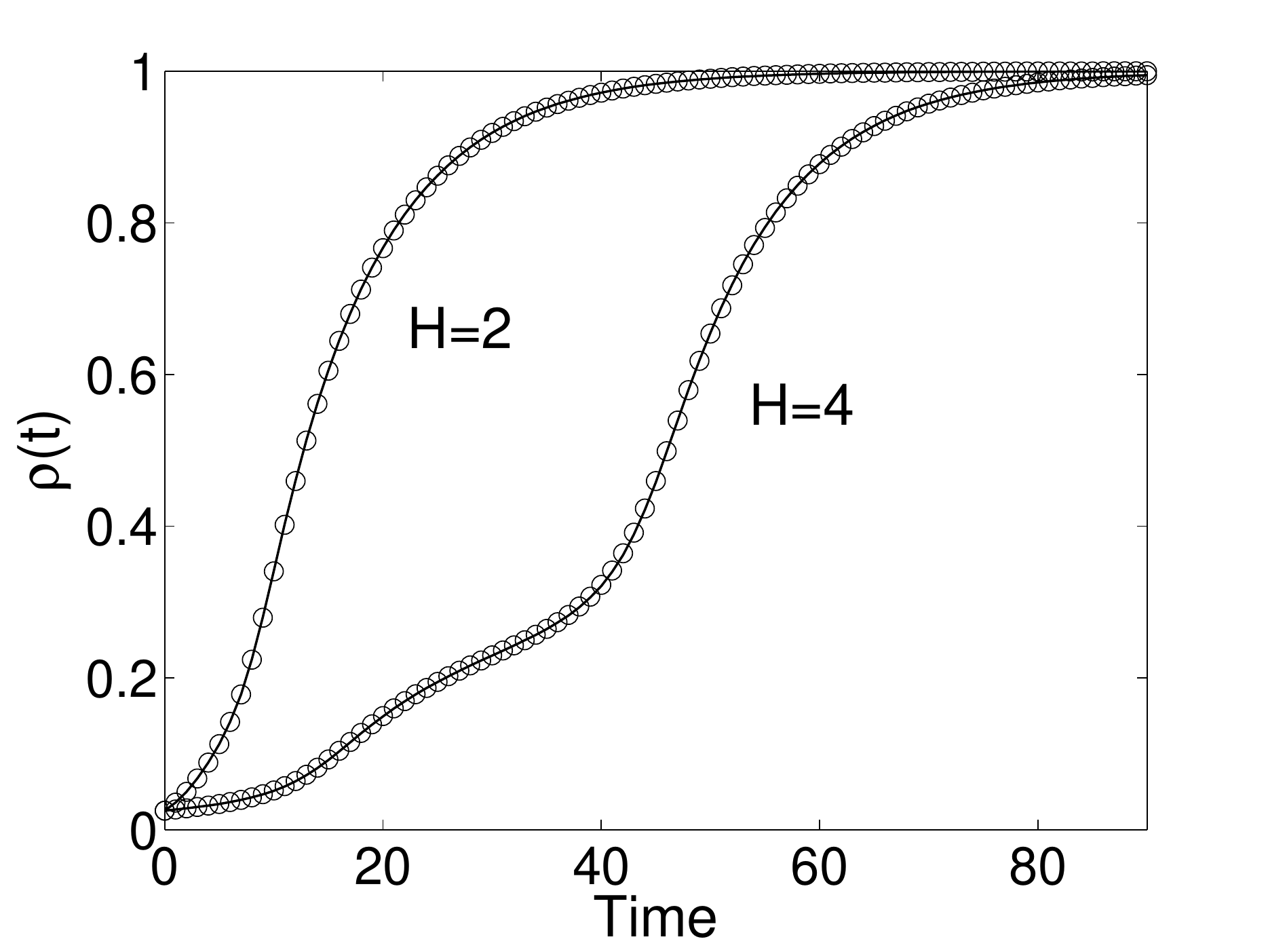}\\(b)\\
\includegraphics[width=0.45\textwidth]{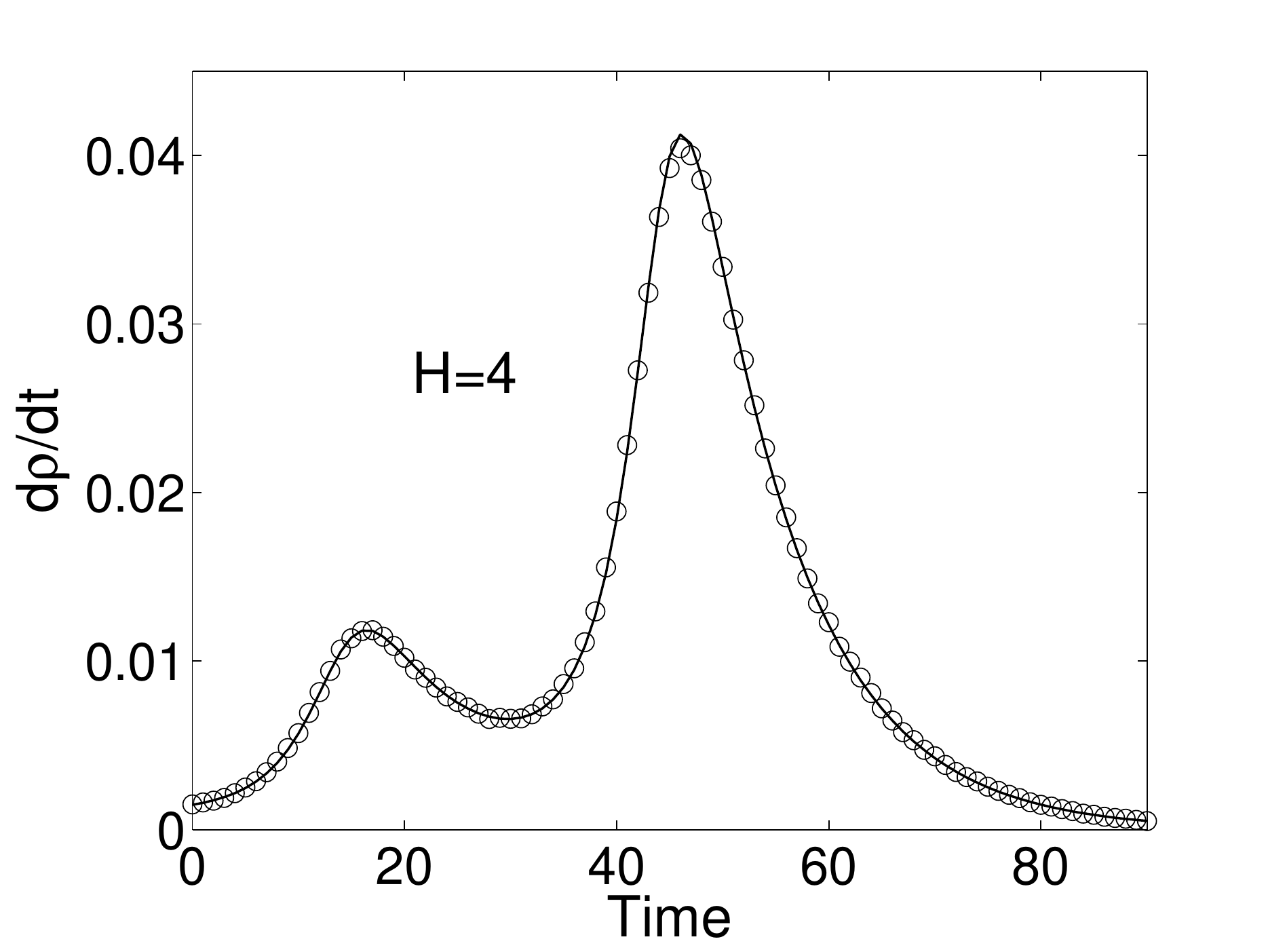}
\end{tabular}
\caption{ Analytical (solid lines) and simulation (circles) results for the activation dynamics. The upper panel shows the fraction of active nodes vs time for threshold parameter $H=2$ and $H=4$. The lower panel shows the activation rate $d\rho/dt$ vs time for $H=4$.}
\label{fig:analysis}
\end{figure}
Equations~\ref{eq:dyn0}  determine the time evolution of the cascading process in each group. Let  $\rho(t) = \alpha \rho_a(t) + (1-\alpha)\rho_{b}(t)$, $\alpha = N_a/(N_a + N_b)$, be the fraction of active nodes in the whole network. In Figure~\ref{fig:analysis}  we compare the solutions obtained from Equations~\ref{eq:dyn0}  with  the results of simulations on randomly generated graphs for the same network parameters  but two different values of the threshold parameter.  The parameters of the network are $N_a=5000$, $N_b=15000$, $z_{aa} =z_{bb}=15 $, $z_{ab} = 4$. The fraction of seed nodes is $\rho_a^0=0.1$, and $\tau^{-1}=0.1$. The simulations are averaged over $100$ random realizations. 

The agreement between the analytical prediction  and results of the simulations is quite good.  The network settles to the same steady state for both values of the threshold parameter $H$: that is, all of the nodes are activated at the end of the cascading process.   However, the transient dynamics depend on the threshold parameter $H$. For $H=2$, activation spreads very quickly through both communities and after a short interval all of the nodes are activate.  For $H=4$, on the other hand, the fraction of active nodes seems to saturate, then, in later iterations, $\rho(t)$  increases rapidly and eventually all the nodes become active.  In Figure~\ref{fig:analysis}(b) we plot the rate of activation process $d\rho/dt$ vs time for $H=4$.  Apparently, the peak rates of activation in the two communities are separated in time.  We call this phenomenon {\it two-tiered dynamics}. We would like to note that previously such a  multi--peak structure has been observed in Ref.~\cite{gupta1989}, where the authors studied the impact of different mixing patterns  on the spread of sexually transmitted infection. 

To better understand how two--tiered dynamics arises, we will examine a simplified scenario. Let us assume that seed nodes are chosen among $a$--nodes only, so that $\rho_b^0=0$. Further, let us assume that the coupling between two populations is not very strong, so that the  cascading process among $a$--nodes is not affected by cross-group links.  Hence, the fraction of active $a$ nodes evolves according to the following equation 
\begin{equation}
\label{eq:dyn2}
\tau \frac{d\rho_{a}}{dt} = - \rho_{a}  + g_a(z_{aa}\rho_a)
\end{equation}
where we have defined 
\begin{equation}
\label{eq:gna}
g_a(x) = 1  - (1 - \rho_{a,b}^0) Q(H, x)
\end{equation}
The fraction of the population that will be activated at the end of the cascading process is determined from the following equation:
\begin{equation}
\label{eq:singlepop}
\rho_a^s=g_a(z_{aa}\rho_a^s)
\end{equation}

\begin{figure}[tbp]
 \center
\includegraphics[width=0.45\textwidth]{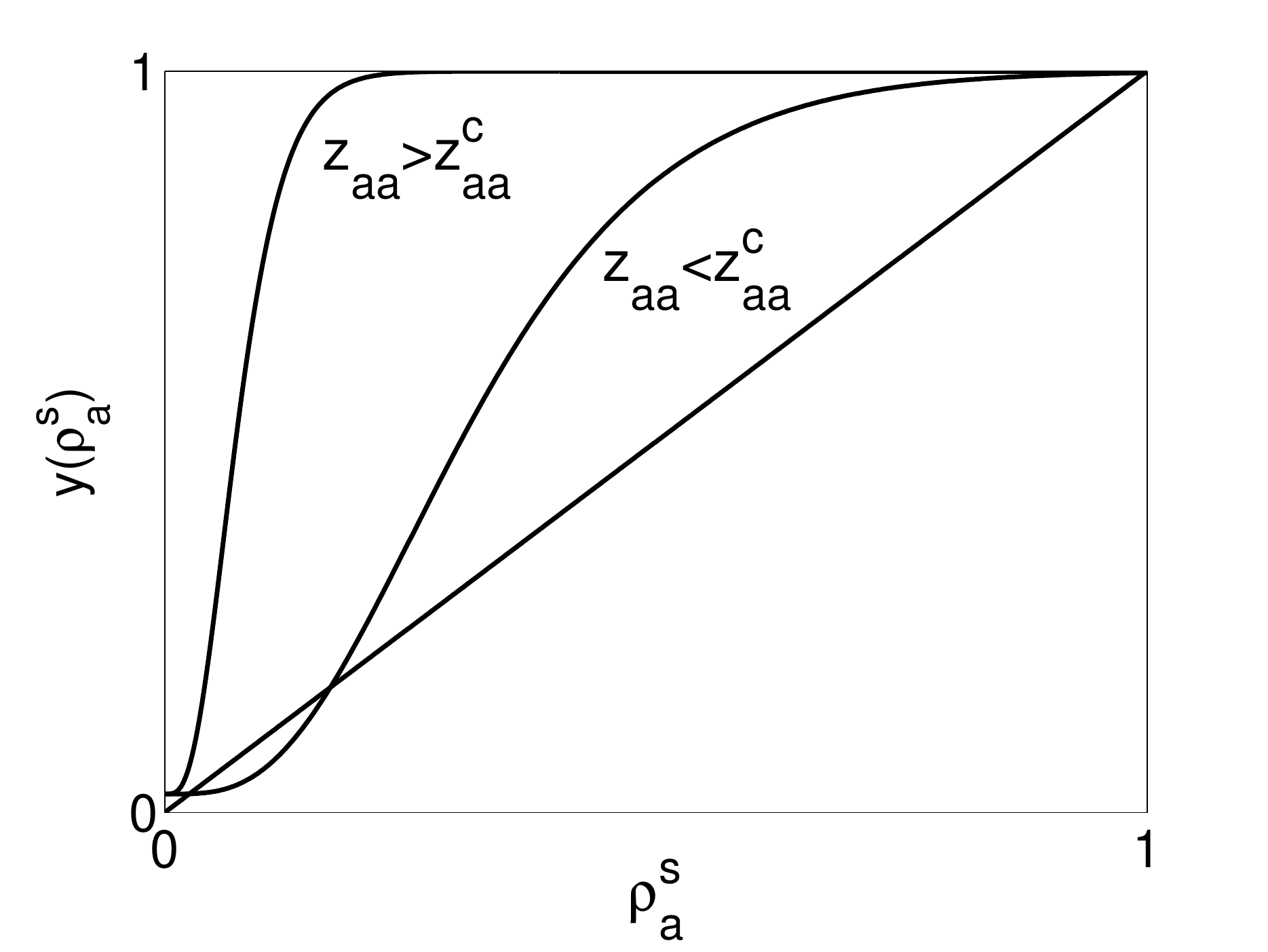} 
\caption{Graphical representation of the Equation~\ref{eq:singlepop}. Plotted are the straight line   $y=\rho_a^s$, and the function $y=g_a(z_{aa}\rho_a^s)$ for two different values of $z_{aa}$}
\label{fig:singlepop}
\end{figure}

Note that for sufficiently dense networks (i.e., the connectivity of all nodes is greater than the threshold $H$) $\rho_a^s=1$ is always a solution. However, it is not always the {\em only} solution. This is shown graphically in Figure~\ref{fig:singlepop}, where we plot both sides of Equation~\ref{eq:singlepop} as a function of $\rho_a^s$ for two different connectivities. For a given fraction of seed nodes the steady--state fraction of active nodes is determined by the connectivity $z_{aa}$. In particular, for sufficiently large values of $z_{aa}$, the only intersection of the curve with the line happens at $\rho \approx 1$,  aside from exponentially small correction of order $\sim z^{H-1}e^{-z_{aa}}$, indicating that the activation will spread globally. If one decreases $z_{aa}$, however, other solutions appear as shown by the two intersections of $\rho_a$ and $g_a(z_{aa}\rho_a)$  in Figure~\ref{fig:singlepop}. Specifically, there is a  critical value $z_{aa}^c$ so that for $z_{aa} < z_{aa}^c$ the cascading dynamics dies out, while for $z_{aa} >z_{aa}^c$ it spreads throughout the system. Let us define $x=z_{aa}\rho_a^s$, and rewrite Equation~\ref{eq:singlepop} as $z_{aa}^{-1}x=g(x)$. At the critical point, the line $z_{aa}^{-1}x$ must be tangential to $g(x)$. It is then straightforward to demonstrate that the critical connectivity is given by 
\begin{equation}
z_{aa}^c = [g_a^{\prime}(x_0)]^{-1}\equiv \biggl [ (1-\rho_a^0)e^{-x_0}\frac{x_0^{H-1}}{(H-1)!} \biggr ] ^{-1}
\label{eq:zc}
\end{equation}
 where $x_0$ satisfies the following equation:
%\begin{equation}
%\label{eq:x0}
%g^{\prime}(x_0) = g(x_0)
%\end{equation}
\begin{equation}
x_0g_a^{\prime}(x_0)=g_a(x_0)
\label{eq:x0}
\end{equation}
\comment{To demonstrate this, let us define $x=z_{aa}\rho_a^s$, and rewrite Equation~\ref{eq:singlepop} as $\frac{1}{z_{aa}}x=g(x)$. The tangential line of $g(x)$ at a point $x=x_0$ is given by $g^{\prime}(x_0)(x-x_0) + g(x_0)$. Requiring the slope of the tangential to equal $1/0z_{aa}$, and the intercept to be zero. }

\begin{figure}[!h]
 \center
\includegraphics[width=0.5\textwidth]{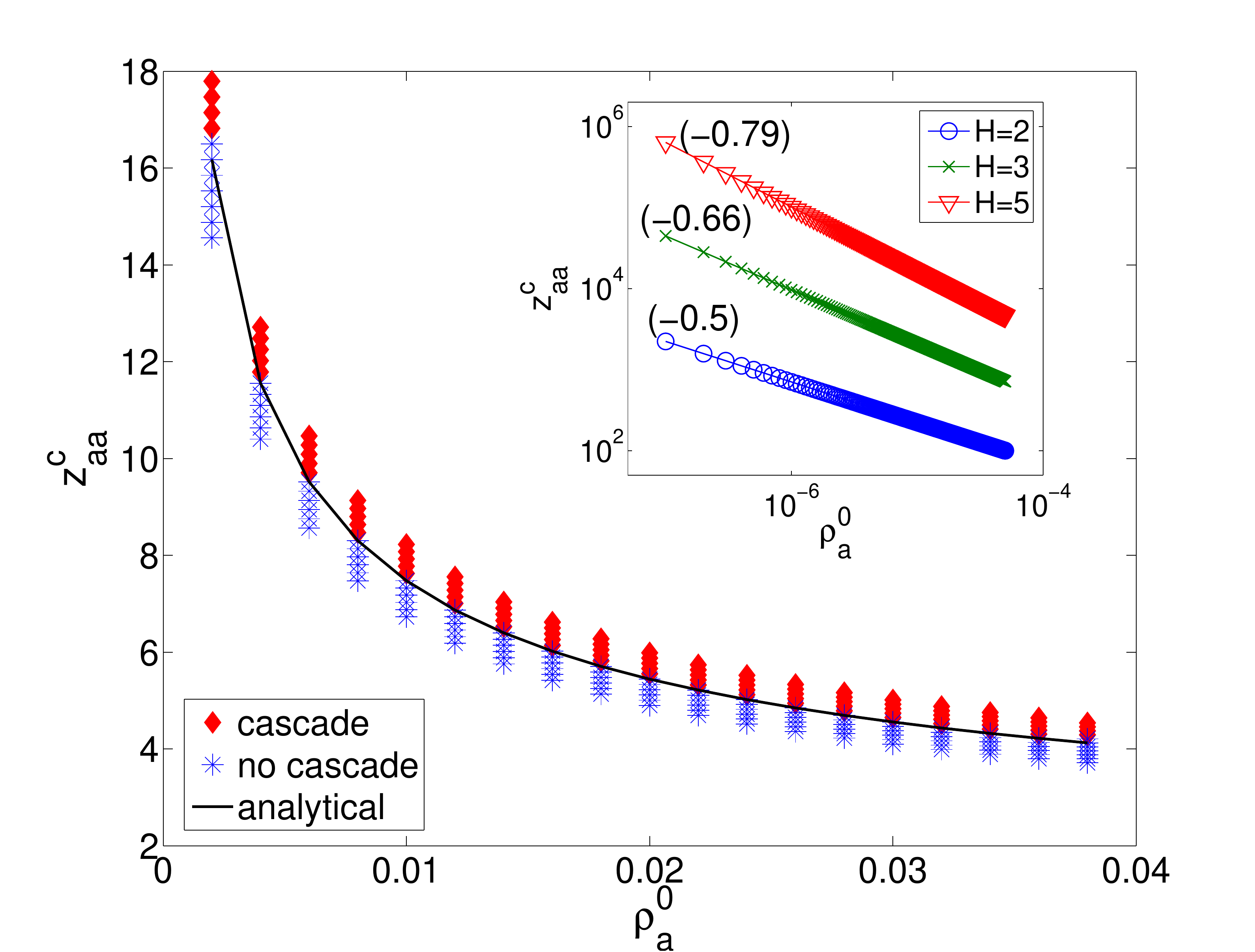} 
\caption{(Color online) The critical connectivity vs fraction of seed nodes for threshold parameter $H=2$. The inset shows the scaling behavior of $z_{aa}^c$ for different $H$ (the numbers in parenthesis are the slopes of corresponding lines).}
\label{fig:zc}
\end{figure}

Equations~\ref{eq:zc} and ~\ref{eq:x0} determine that critical connectivity needed to cause a global cascade among $a$ nodes for a given fraction of seed nodes and the threshold parameter $H$. In Figure~\ref{fig:zc} we compare the analytical prediction with simulation results for $H=2$. The simulations were done for a graph with $5\times 10^4$ nodes, and for $100$ random trials. Each parameter pair $(\rho_a^0, z_{aa})$ was considered to be above the critical line if a global cascade was observed in the majority of trials  for that parameters. Again, the agreement of analytical prediction and the simulation results are excellent. 

Let us examine the behavior of the critical connectivity in the limit of small $\rho_a^0$. The equation~\ref{eq:x0} can be rewritten as
\begin{equation}
e^{-x_0} \biggl ( \frac{x_0^H}{(H-1)!} + \sum_{k=0}^{H-1}\frac{x_0^k}{k!} \biggr)  = \frac{\rho_0^a}{1-\rho_0^a}
 \end{equation}
Assuming $\rho_a^0, x_0 \ll 1$ we obtain in the leading order
\begin{equation}
x_0 \approx  \biggl [ \frac{H!}{H-1}  \rho_a^0 \biggr ]^{\frac{1}{H}}
\label{eq:x02}
\end{equation}
Finally, using Equation~\ref{eq:zc} we obtain the following scaling behavior
\begin{equation}
z_{aa}^c \propto (\rho_a^0)^{-\frac{H-1}{H}}.
\label{eq:scaling}
\end{equation}
which is demonstrated in the inset of Figure~\ref{fig:zc}. We also note that at the critical point the convergence time diverges as $T_{conv}\propto (z-z_{aa}^c)^{-1/2}$.

Now consider the cascading dynamics in the second group. Initially, there are no active nodes in this group. As more and more $a$ nodes are activated, the activation will spread to the $b$ nodes for sufficiently large across--group connectivity $z_{ba}$.   The activation dynamics is again governed by  an equation similar to the Equation.~\ref{eq:dyn2}. In particular, the steady state fraction of active $b$ nodes  satisfies the following equation:
\begin{equation}
\label{eq:steadyb}
\rho_b^0 = 1  -  Q(H, z_{bb}\rho_b^0 + \lambda) \equiv g_b(z_{bb}\rho_b^0 + \lambda). 
\end{equation}
where  $\lambda = z_{ba}\rho_a^0$. Clearly, if $\lambda$ is sufficiently large, then the cascade will propagate among $b$ nodes independent of the within--group connectivity $z_{bb}$. And vise versa, however large the connectivity $z_{bb}$,  there is a critical value of $\lambda_a^c$ so that for $\lambda < \lambda^c$ there will be no cascade among the  $b$ nodes. Let us define $x=z_{bb}\rho_b^0 + \lambda$ and rewrite the steady state equation as follows: 
\begin{equation}
\frac{x-\lambda}{z_{bb}} = g_b(x)
\end{equation}
Using the same reasoning as for the $a$ nodes, it is easy to show that the critical point  is given by 
\begin{equation}
\lambda^c = x_0 - z_{bb}g_b(x_0)
\label{eq:lambda}
\end{equation}
where $x_0$ is the smaller of the roots of the following equation:
\begin{equation}
g_b^{\prime}(x_0) = \frac{1}{z_{bb}}
\label{eq:x0b}
\end{equation}
Note that for $\rho_a^0=1$ $\lambda_c$ is simply  the critical across--group connectivity $z_{ab}^c(z_{bb})$ for which the cascade will spread to $b$ nodes, assuming that all $a$ nodes have already been activated.  Hence, equations~\ref{eq:lambda} and~\ref{eq:x0b}  implicitly define a critical line $z_{bb}^c(z_{ba})$ on the $z_{bb}-z_{ba}$ plane.  Note that on this critical line the convergence time of the cascading process among the $b$-nodes,  and consequently the separation of two activity peaks, is infinite. For a fixed within--group connectivity $z_{bb}$ the two--tiered structure will be present provided that $z_{ba}$ is only slightly above the critical line. To be more precise, let $\rho_a^{max}$ be the fraction of active $a$ nodes that corresponds to the maximum activation rate among $a$ nodes. This can be found from Equation~\ref{eq:dyn2} by differentiating the right hand side with respect to $\rho_a$ and setting it to zero, which yields $z_{aa}g_a^{\prime}(z_{aa}\rho_a^{max})=1$. If the across---group connectivity is smaller than $\lambda^c/\rho_a^{max}$, then the cascade will not spread to $b$--nodes until the rate of activation spreading among $a$ nodes   starts to decline from its peak. Consequently, the two--tiered pattern will be present for the range $\lambda^c < z_{ba} < \lambda^c /\rho_a^{max}$.

\section{Summary}
To summarize, we have considered a simple cascading model on a random network consisting of two--loosely coupled communities. For a sufficiently weak coupling between two communities the dynamics of the activity spreading demonstrates two--tiered structure, that is,  the peak rates of the cascading processes in two communities are separated in time. This pattern is reminiscent of multi--peak structure of sexually transmitted infection dynamics, previously reported in Ref.~\cite{gupta1989}.  We studied this phenomenon  both experimentally  and theoretically using annealed approximation, and obtained a good agreement between analytical results and  simulations.  Although our model is for undirected binary graphs, generalizations to directed and/or weighted graphs is straightforward. Directed models can be relevant if the interactions between two nodes are not symmetric. 

The results presented here might have  implications in problems such as epidemiology, viral marketing, and so on. Consider, for example, the problem of minimizing the number of seed nodes that will cause  a global cascade in a given network, or more generally, the problem of maximizing a certain utility function $f(N_0, N^s)$, where $N_0$ is the number of seed nodes, and $N_s$ is the expected size of the cascade. Our results suggest that  simple strategies that are suitable for homogenous networks (e.g., choosing  nodes with high connectivity, or at random), might lead to a  sub--optimal solution for networks with strongly modular structure. We note, however, that in order to assess the implication of our findings in real world problems, one needs to generalize the approach developed here to more complex networks.

\bibliographystyle{apsrev}
%\bibliography{cascade2}

\end{document}